\documentclass[sigconf]{acmart}

\usepackage{sections/macros}
\usepackage{amsmath,amsfonts}
\usepackage[ruled,linesnumbered]{algorithm2e}
\usepackage{graphicx}
\usepackage{textcomp}
\usepackage{url}
\usepackage{xcolor}
\usepackage{enumitem}
\usepackage{threeparttable}
\usepackage{booktabs}
\usepackage{caption}         
\usepackage{subcaption}      

\setcopyright{none}
\begin{document}

\title{Lagom: Unleashing the Power of Communication and Computation Overlapping for Distributed LLM Training
}


\author{Guanbin Xu$^{\dagger}$,
ZhenGuo Xu$^{\dagger}$,
Yuzhe Li$^{\dagger}$,\\
Youhui Bai$^{\dagger}$,
Ping Gong$^{\dagger}$,
Chaoyi Ruan$^{\ddagger}$\\
Cheng Li$^{\dagger,\circ}$
}
\affiliation{
$^{\dagger}$University of Science and Technology of China
$^{\circ}$ Anhui Province Key Laboratory of Biomedical Imaging and Intelligent Processing, \\
Institute of Artificial Intelligence, Hefei Comprehensive National Science Center
\country{China}
}

\affiliation{
$^{\ddagger}$National University of Singapore
\country{Singapore}
}
\begin{abstract}

Overlapping communication with computation is crucial for distributed large-model training, yet optimizing it—especially when computation becomes the bottleneck—remains challenging.
We present \pn, a system that co-tunes communication parameters to balance resource usage between computation and communication.
By introducing a unified cost model and a priority-based search algorithm, \pn reduces optimization complexity from exponential to linear.
Evaluations on high- and low-bandwidth GPU clusters show that \pn achieves 1.07-1.33$\times$ and 1.03-1.27$\times$ speedup over NCCL and AutoCCL across diverse models and parallelizations.
\end{abstract}
\maketitle
\section{Introduction}
\label{sec:intro}
The increasing scale of Deep Neural Networks (DNNs) is driving the need for distributed training, which is often limited by communication bottlenecks.
To this end, a common optimization is to overlap communication with computation~\cite{peng2019generic, cheng2025concerto, chen2024centauri}.
However, achieving optimal performance of overlap remains challenging, as communication operations contend with computation kernels for limited GPU resources.
In practice, mature overlap strategies in industry often hide most communication latency~\cite{jiang2024megascale}, shifting the primary bottleneck to computation. Yet, even in such computation‑bottlenecked scenarios, communication still interferes with computation, leaving overlooked opportunities for further optimization.

Existing approaches to managing contention have notable limitations. Liger~\cite{du2024liger, deepep2025} mitigates contention by statically limiting communication's GPU resources, but its fixed allocation lacks adaptability. 
Libra~\cite{liu2023libra} models overlap cost for GPU thread selection in data parallelism, yet it ignores key parameters (\eg chunk size and number of channels) that we find have a greater impact on computation (Sec.~\ref{design:contention_modeling}). 
AutoCCL~\cite{xu2025autoccl}, the state-of-the-art communication tuner, addresses parameter tuning and is effective when communication is the bottleneck.
While these approaches address specific aspects of overlap, they fail to holistically manage the dual-bottleneck scenario, especially when computation becomes the bottleneck where our analysis reveals that communication contention still degrades the performance of the bottlenecked computation by up to 35\%.

\begin{figure}[!htbp] 
    \centering
    \includegraphics[width=1\linewidth]{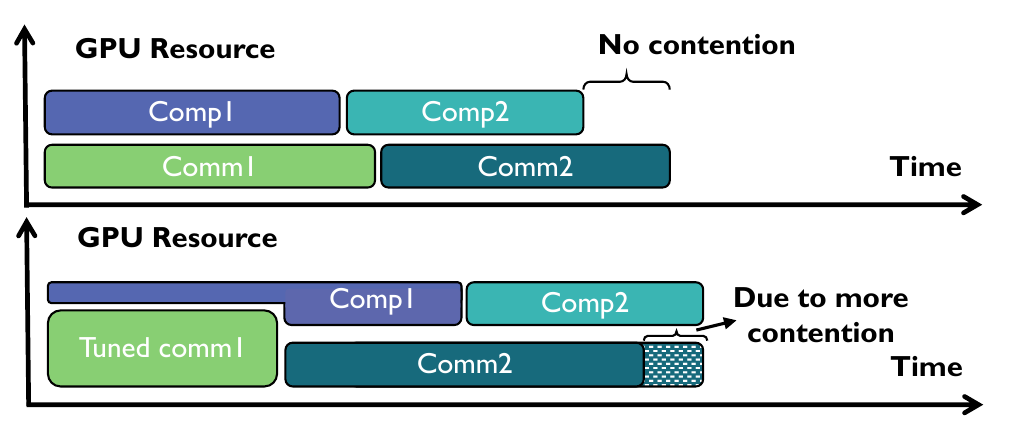} 
    \caption{Side effects from tuning a single communication. Top: baseline execution. Bottom: after tuning Comm1, increased resource contention delays Comp2. }
    \label{fig:multi_comm_comp} 
\end{figure}

Our analysis reveals that by strategically allocating fewer resources to communication, computation can utilize more resources, thereby accelerating the entire overlap. 
However, achieving this efficient overlap through parameter tuning is nontrivial, facing key challenges:
(1) \textbf{Contention Complexity}: The relationship between the vast tunable parameter space and resource contention during the overlap of communication and computation operators—which must share multiple GPU resources—remains an open and poorly understood research problem. 
(2) \textbf{Exponential Search Space}: To prevent deadlock, collective communications are serialized, creating a chain of temporal dependencies.
Consequently, tuning parameters for one communication (e.g., Comm1) affects not only its own execution but also cascades to delay dependent computations and other communications, as visualized in Fig.~\ref{fig:multi_comm_comp}. 
This interdependence makes the joint optimization space grow exponentially with the number of communications—i.e., exceeding $1,000,000^N$ for N communications.
An exhaustive search is infeasible, as it would require hundreds of GPU hours even for moderate N.

This paper presents \pn, a novel system for tuning collective communication parameters. By introducing an overlap performance model and a resource-efficiency search heuristic, \pn efficiently identifies near-optimal configurations for overlapping communication and computation. The main contributions are:

\noindent \textbf{(1) Unified Overlap Performance Model}: We formalize a unified cost model for computation-communication overlap that captures performance in both computation-bound and communication-bound regimes, establishing a foundational basis for optimization.

\noindent  \textbf{(2) Contention Analysis and Modeling}: We analyze and categorize contention into two main types—SM competition and global resource contention—and model their relationship with tunable communication parameters.

\noindent  \textbf{(3) Tuning Algorithm and Evaluation}: We introduce a fast search algorithm based on model-guided heuristics that reduces the optimization complexity and achieves 1.07-1.33$\times$ speedup across diverse models and parallelisms. The code for \pn will be open-sourced soon.

\section{Background and Motivation}
\label{sec:background}

\subsection{Parallelism and Overlapping}
\begin{figure}[thbp] 
    \centering
    \includegraphics[width=\linewidth]{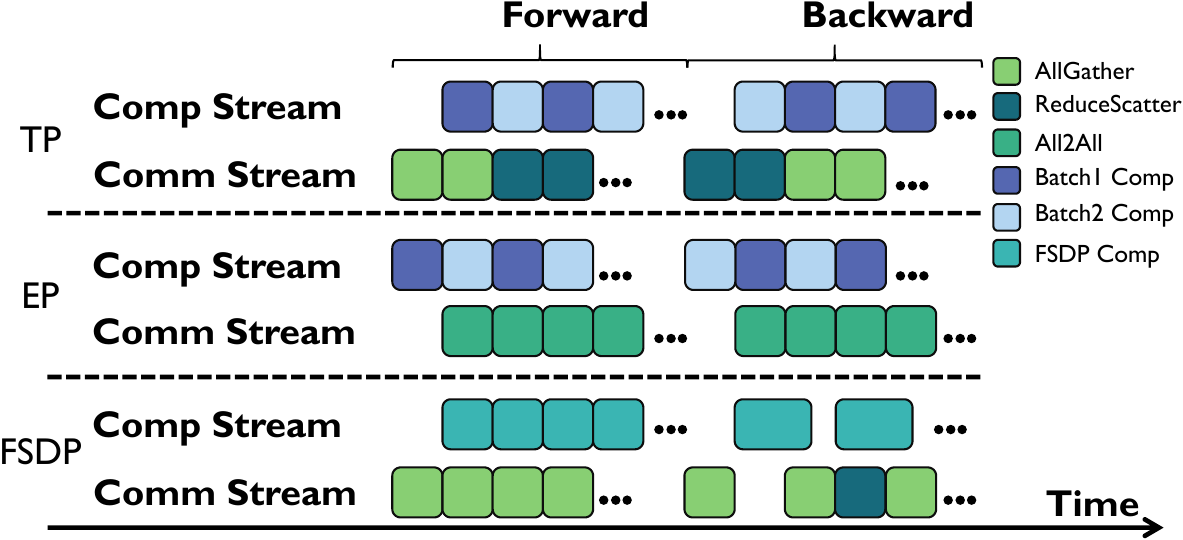} 
    \caption{Overlaps of Different Parallelism}
    \label{fig:Parallelism overlap} 
\end{figure}

As data and model scales continue to grow, distributed training has become essential~\cite{fan2021dapple, unger2022unity}. 
Modern systems adopt various parallelisms that exploit communication–computation overlap as Fig.~\ref{fig:Parallelism overlap}. The main forms include:

\noindent\textbf{Tensor Parallelism (TP)}~\cite{shoeybi2019megatron} partitions large operators across devices to ease memory pressure, but incurs synchronization overhead (e.g., \texttt{AllReduce}). Domino~\cite{wang2024domino} mitigates this via batch-level pipelining, though perfect overlap remains difficult.

\noindent\textbf{Expert Parallelism (EP)}~\cite{shazeer2017outrageously} distributes experts across devices and relies on frequent \texttt{AlltoAll}.
While efforts like the dual batch method aim to overlap communication with computation by resource partitioning~\cite{liu2024deepseek}, the cluster-specific resource allocation and AlltoAll-only implementation limit general applicability.

\noindent\textbf{Fully Sharded Data Parallelism (FSDP)} (\cite{zhao2023pytorch, rajbhandari2020zero}) exhibits a more intricate overlap pattern, where each layer’s computation interleaves with multiple communication phases (\texttt{AllGather} and \texttt{ReduceScatter}) across layers. 
Such multi-stage coupling makes concurrent optimization between computation and communication highly complex, often hindering ideal overlap.

These parallelisms are often combined with data parallelism~\cite{lepikhin2020gshard,shazeer2018mesh,wang2019supporting} and pipeline parallelism~\cite{huang2019gpipe,narayanan2019pipedream} to support even larger-scale training~\cite{zheng2022alpa,jia2020whale,narayanan2021efficient,jiang2024megascale}.

\subsection{Operators in Overlap and Parameter Tuning}

Overlap involves over 1,500 computational operators~\cite{zhao2023pytorch}, primarily implemented in closed-source libraries like cuBLAS~\cite{cublas} and cuDNN~\cite{cudnn}, making direct modification infeasible. 
In contrast, collective communications benefit from optimized open-source libraries like NCCL~\cite{nccl}. AutoCCL~\cite{xu2025autoccl} identified six key parameters (Algorithm, Protocol, Transport, Number of Channels, Number of Threads, Chunk Size) that govern communication performance. 
They make a key observation that the configuration space can be addressed using a divide-and-conquer strategy. It first partitions the space into subspaces based on implementation-related parameters (Algorithm, Protocol, Transport), then models resource-related parameters (Number of Channels, Number of Threads, Chunk Size) within each subspace. 
Crucially, instead of explicitly modeling complex interference patterns, AutoCCL employs online sampling with real-time performance feedback to efficiently search resource-related parameters in each subspace. \textbf{We extend this finding by showing that these parameters also regulate GPU resource allocation during overlap, thereby critically affecting communication-computation contention.}

\subsection{Challenges}
However, the interaction between communication and computation is inherently difficult.
The dynamic nature of multi-communication tuning expands the search space exponentially, making joint optimization challenging.
These challenges mainly arise from: 
(1) \textbf{Complex Contention Between Communication and Computation.}
    Communication operations share core GPU resources such as Streaming Multiprocessors (SMs), global memory bandwidth, and L2 cache with computations.
    The degree of contention varies with communication type, message type, concurrency, and hardware design, making performance modeling highly nonlinear.
(2) \textbf{Exponential Growth of the Search Space for Multi-Communications.} 
    In overlapping scenarios, multiple communications interfere with each other and with computation.
    For instance, tuning the parameters of one communication operation may change its completion time, which, in turn, affects the overlap between subsequent communications and computation, thereby altering the overall system performance.
    With $ n $ communication operations each having $ k $ configuration options, the joint search space grows as $ k^n $, which calls for an effective reduction in dimensionality.

\section{Design}
\label{sec:design}

\subsection{Problem Definition}

\begin{table}[ht]
    \centering
    \caption{Key notations used in the cost model.}
    \label{tab:notation}
    \resizebox{0.97\columnwidth}{!}{
    \begin{tabular}{ll}
    \toprule
    \textbf{Notation} & \textbf{Description} \\
    \midrule
    $X, Y, Z$ & Communication, computation, and total time. \\
    $s_j$ & \(j\)-th communication config (A, P, T, NC, NT, C). \\
    $TB_i$ & \(i\)-th computation's threadblocks. \\
    $\lambda$ & Total number of SMs. \\
    $D_i$ & Data size per thread block in the \(i\)-th computation. \\
    $\overline{B}$ & Peak global memory bandwidth. \\
    $V_j$ & Global bandwidth of the \(j\)-th communication. \\
    \bottomrule
    \end{tabular}
    }
\end{table}


\begin{equation}
\label{costmodel:overlap}
\min_{\{s_1,...s_N\}} Z =\max \left( Y,X \right)= \max \left( \sum_{i=1}^{M} y_i, \ \sum_{j=1}^{N} x_j^{s_j} \right) 
\end{equation}
To address these challenges, we now introduce a unified cost model and analyze its behavior across different bottleneck scenarios in computation-communication overlap.
For an overlap, let it contain $M$ computation operators and $N$ communication operators. 
We denote the execution time of the i-th computation operator as $y_i$, and the execution time of the j-th communication operator under configuration $s_j$ as $x_j^{s_j}$.
With computations and communications serialized on separate CUDA streams, the total computation time is $Y=\sum_{i=1}^{M}y_i$, the total communication time is $X=\sum_{j=1}^{N}x_j^{s_j}$, and the overall makespan is given by $Z=max(X, Y)$.
As shown in Equation \eqref{costmodel:overlap}, 
the optimization objective is to minimize the overall makespan $Z$ by selecting optimal configurations $s_j$ for all communication operators.
We now analyze the two bottleneck scenarios separately. 

\begin{figure}[!htbp]
\centering
\begin{align}
\min_{\{s_1,...s_N\}} Z = \min_{\{s_1...s_N\}} \sum_{j=1}^{N} x_j^{s_j} = \sum_{j=1}^{N} \min_{s_j} x_j^{s_j}
\label{costmodel:comm_dominant}
\end{align}
\end{figure}

\noindent\textbf{Communication Bottleneck.}
First, in the communication-bound case described by Equation~\ref{costmodel:comm_dominant}, the overlap makespan simplifies to the cumulative time of all communications. 
The key challenges are: (1) modeling how configuration $s_j$ affects communication time $x_j^{s_j}$, and (2) capturing the impact of computational contention on communication.
AutoCCL~\cite{xu2025autoccl} addresses this by employing a subspace coordinate descent method for search, and leverages online sampling with real-time feedback to avoid explicit modeling of the contention.

\begin{equation}
\label{costmodel:comp_dominant}
\min_{\{s_1...s_N\}} Z = \sum_{i=1}^{M} y_i 
\end{equation}

\noindent\textbf{Computation Bottleneck.}
Symmetrically, Equation~\ref{costmodel:comp_dominant} describes the computation-bound scenario, where the overlap makespan is determined by the sum of computation operator times.
Two key challenges arise in this case: (1) Communication parameters influence computational performance by governing how communications utilize shared resources (further discussed in Sec.~\ref{design:contention_modeling}); 
(2) Due to the serialized execution of communications, tuning an earlier communication changes the start time and overlap window of subsequent operators, creating a cascade effect on the computation. This requires considering the entire configuration set ${s_1, s_2, \dots, s_N}$ jointly. With the configuration space per communication exceeding $r=10^6$ options~\cite{xu2025autoccl}, the joint search space becomes $O(r^N)$ for $N$ communication operators, which is computationally intractable for practical values of $N$.

We now analyze how to tune communication parameters in the computation-bound overlap scenario.

\subsection{Contention Modeling}
\label{design:contention_modeling}
\begin{figure}[!thbp]
    \centering
    \begin{minipage}[b]{0.55\linewidth}  
        \centering
        \includegraphics[width=\linewidth]{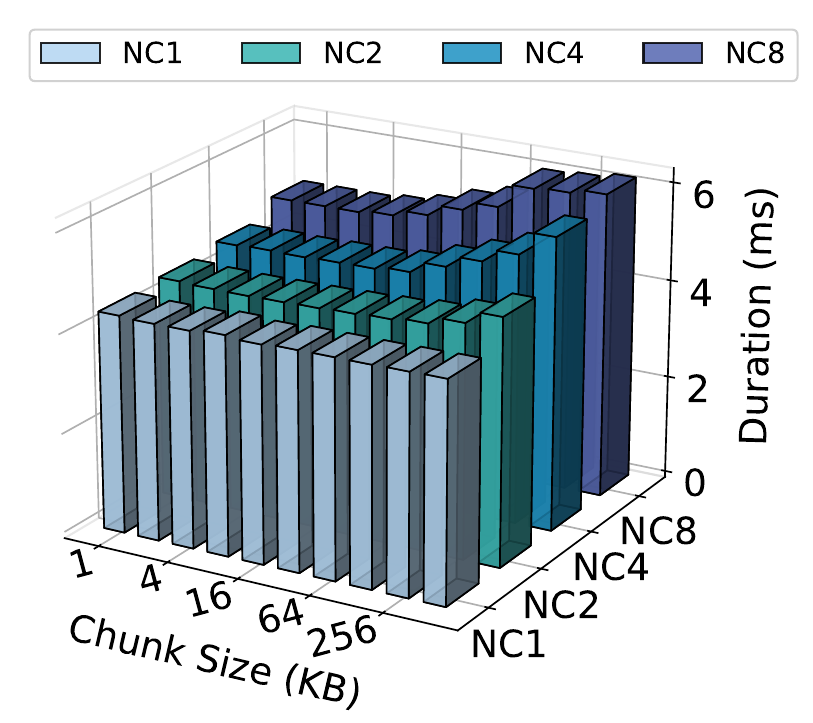}
        \subcaption{Various $NC$ and $C$}
        \label{various_nc_c}
    \end{minipage} 
    \hfill
    \begin{minipage}[b]{0.4\linewidth}  
        \centering
        \includegraphics[width=\linewidth]{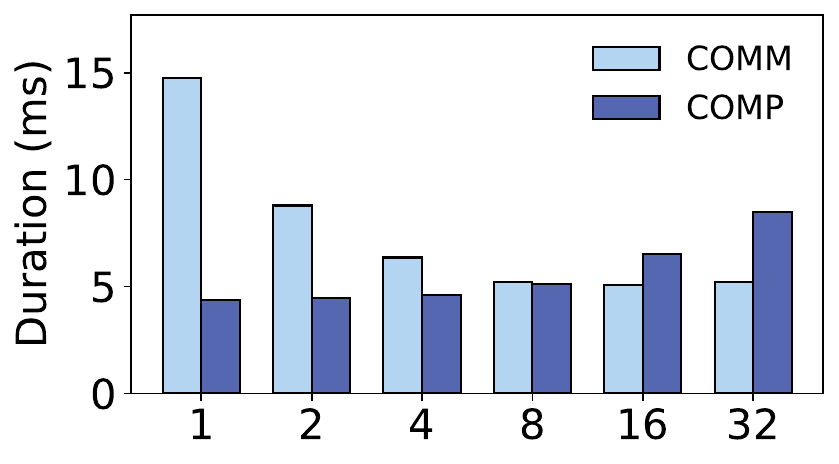}
        \subcaption{Various $NC$}
        \label{various_nc}
        \vspace{0.2cm}  
        \includegraphics[width=\linewidth]{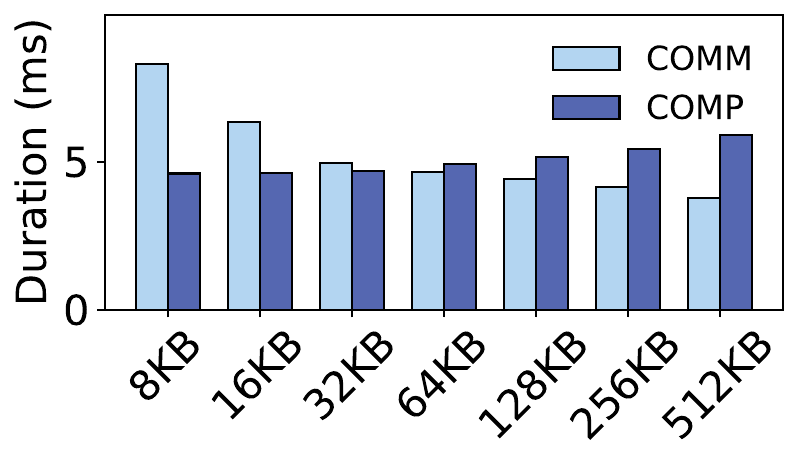}
        \subcaption{Various $C$}
        \label{various_c}
    \end{minipage}
    \caption{Duration of FFN when AllReduce(32MB) with various NC and C (NT=128) on 8 $\times$ A40pcie}
    \label{fig.3.various_nc_c}
\end{figure}
Building on AutoCCL~\cite{xu2025autoccl}, we adopt a divide-and-conquer strategy based on implementation-related parameters (Algorithm, Protocol, Transport) before analyzing the effect of resource-related parameters. 
Our key finding is that while different communication configurations can achieve similar communication performance, they exhibit significant variation in how much they slow down computation.
As shown in Fig.~\ref{fig.3.various_nc_c}, when using NC=16 versus NC=32, communication performance remains nearly identical, but computation time differs by 30.2\%.

We conducted separate tests on NC, NT, and C, and found that NC and C have a significant impact on computation time, while the effect of NT is negligible.
Fig.~\ref{various_nc_c} shows the overall impact of different NC and C configurations on computation time, measured by concurrently running a 32 MB AllReduce and an FFN operator on 8 NVIDIA A40 GPUs with PCIe interconnect.
The results indicate that increasing either NC or C progressively prolongs computation time.
To isolate individual factors, Fig.~\ref{various_nc} shows how communication and computation times vary as NC increases while $C=16 KB$.
Similarly, Fig.~\ref{various_c} shows the effect of increasing C when $NC=4$.
The figures confirm that while larger NC and C values initially reduce communication time (with diminishing returns or even slight increases at larger values), they also significantly slow down computation.
These findings were reproduced consistently on another cluster.

\begin{figure}[!htbp]
\centering
\begin{minipage}{0.46\columnwidth}
    \centering
    \includegraphics[width=\linewidth]{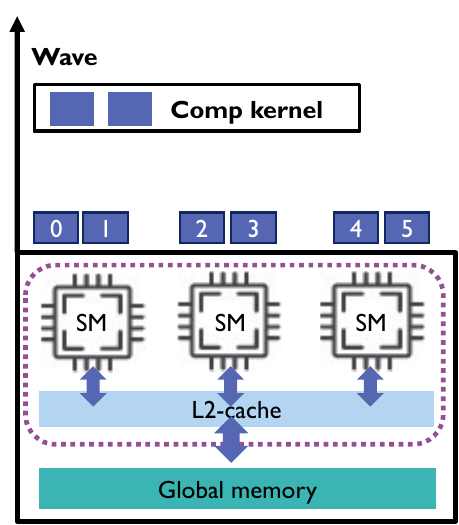}
    \subcaption{Computation Only}
    \label{fig:contention:no}
\end{minipage}
\hfill
\begin{minipage}{0.48\columnwidth}
    \centering
    \includegraphics[width=\linewidth]{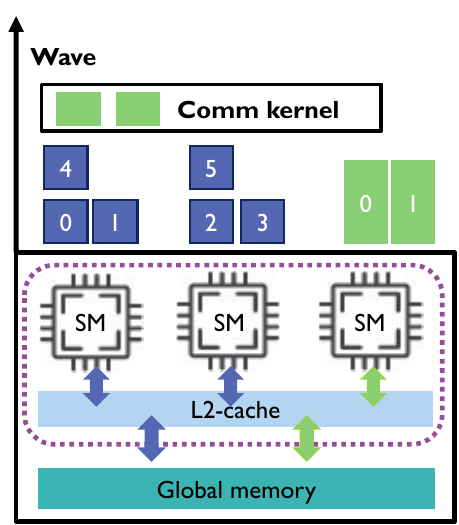}
    \subcaption{Communication Contention}
    \label{fig:contention:yes}
\end{minipage}
\caption{SM and global resource contention.}
\label{fig:contention}
\end{figure}

\noindent\textbf{Cost Model}
We characterize communication-computation contention along two dimensions: SM competition and global resource competition.
As illustrated in Fig.~\ref{fig:contention}, SM competition arises from shared processor occupancy, while global resource contention stems from bandwidth saturation.
\textbf{SM Competition.}
Communication operators occupy a portion of SMs, reducing the resources available for computation and increasing the number of computation waves. 
The assignment of threadblocks, whether for computation or communication, to an SM is constrained by the availability of threads, shared memory, and registers.
Due to this multi-constraint bottleneck, NT has a negligible impact on SM competition.
NC is the dominant factor governing SM occupancy.
\textbf{Global Resource Competition.} Communication also contends for shared global resources such as memory bandwidth and cache.
Each communication operator launches NC threadblocks that read/write data at a granularity of C.
Memory transactions from different threads within a threadblock are coalesced, making NT's impact on global contention relatively minor.
In contrast, larger values of NC and C generate more intensive contention due to increased concurrent data movement.

\begin{align}
\label{costmodel:comp_time}
y_{i} &= \sum_{j=l}^k f_{ij} \times g_{ij} \\
\label{costmodel:comp_wave_num}
g_{ij} &= \lceil \frac{\mu_i}{(\lambda - NC_j)\times TB_i} \rceil \\
\label{costmodel:comp_wave_time}
f_{ij} &= \theta_{ij} + \frac{(\lambda - NC_j) \times TB_i \times D_i}{\overline{B} - V(NC_j, C_j)}
\end{align}
Based on these observations, we build a cost model of the computation $i$ overlapped with communication $l-k$.
Eq.~\eqref{costmodel:comp_time} decomposes computation time $y_i$ into wave count $g_{ij}$ and per-wave latency $f_{ij}$ that is overlapped with the j-th communication.
Eq.\eqref{costmodel:comp_wave_num} shows the number of computation waves $g_{ij}$ is determined by dividing the total required thread blocks $\mu_i$ by the number of available threadblocks per SM after allocating $NC_j$ SMs to communication.
Meanwhile, Eq. \eqref{costmodel:comp_wave_time} shows $f_{ij}$ consists of computation time $\theta_{ij}$ and data transfer time, where the latter depends on the data volume $D_i$ and the effective bandwidth reduced by global resource contention $V(NC_j, C_j)$.  
Notation definitions are summarized in Table~\ref{tab:notation}.
A higher NC occupies more SMs, reducing the resources available for computation and consequently increasing the number of waves \(g_i\), while also imposing additional demand on bandwidth.
A larger C, in turn, amplifies contention for global bandwidth, thereby extending the per‑wave execution time \(f_i\).
Fig.~\ref{fig.3.various_nc_c} empirically validates our model: increasing NC/C beyond optimal points (e.g., NC>8) drastically prolongs computation time despite marginal communication gains.

\subsection{Priority Metric}
When multiple communication and computation tasks execute concurrently, the central challenge lies in efficiently identifying near-optimal configurations from a combinatorial space that scales as $O(1,000,000^n)$. 
Brute-force searching this space is infeasible, necessitating the use of heuristic methods to find high-quality solutions within practical time constraints.
A naive strategy is to optimize communications sequentially. However, this approach fails to account for critical interdependencies: tuning one communication alters its execution timeline, which cascades to affect both the communication and computation phases of subsequent tasks. These temporal and resource-contentions introduce nonlinear interactions that render sequential optimization highly suboptimal.
To illustrate this, an experiment was conducted on A40 GPUs with 2 AllReduce and 7 MatMul executed concurrently.
We adjust $NC$ of one communication at a time to analyze how it impacts overall performance.

\begin{figure}[!htbp]
    \centering
    \begin{minipage}[b]{0.48\linewidth}  
        \centering
        \includegraphics[width=\linewidth]{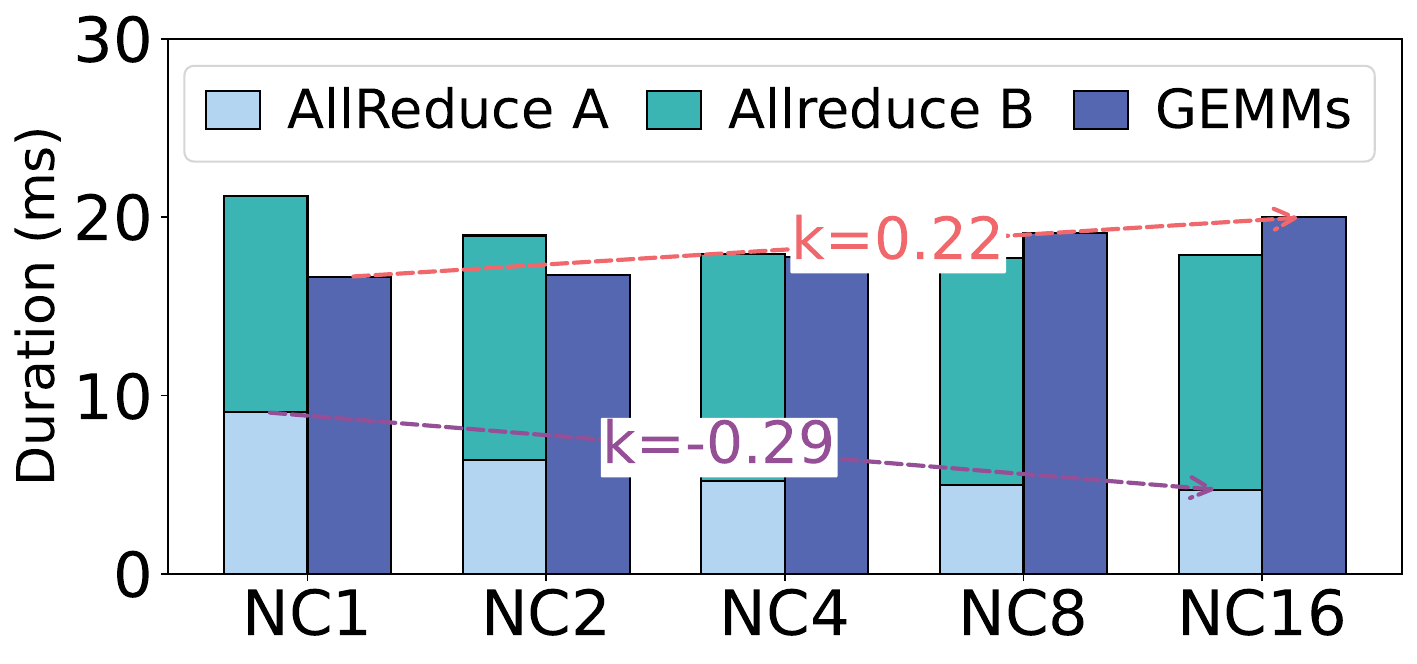}
        \subcaption{tuning AllReduce A only}
        \label{fig:tune.A}
    \end{minipage}
    \hfill
    \begin{minipage}[b]{0.48\linewidth}  
        \centering
        \includegraphics[width=\linewidth]{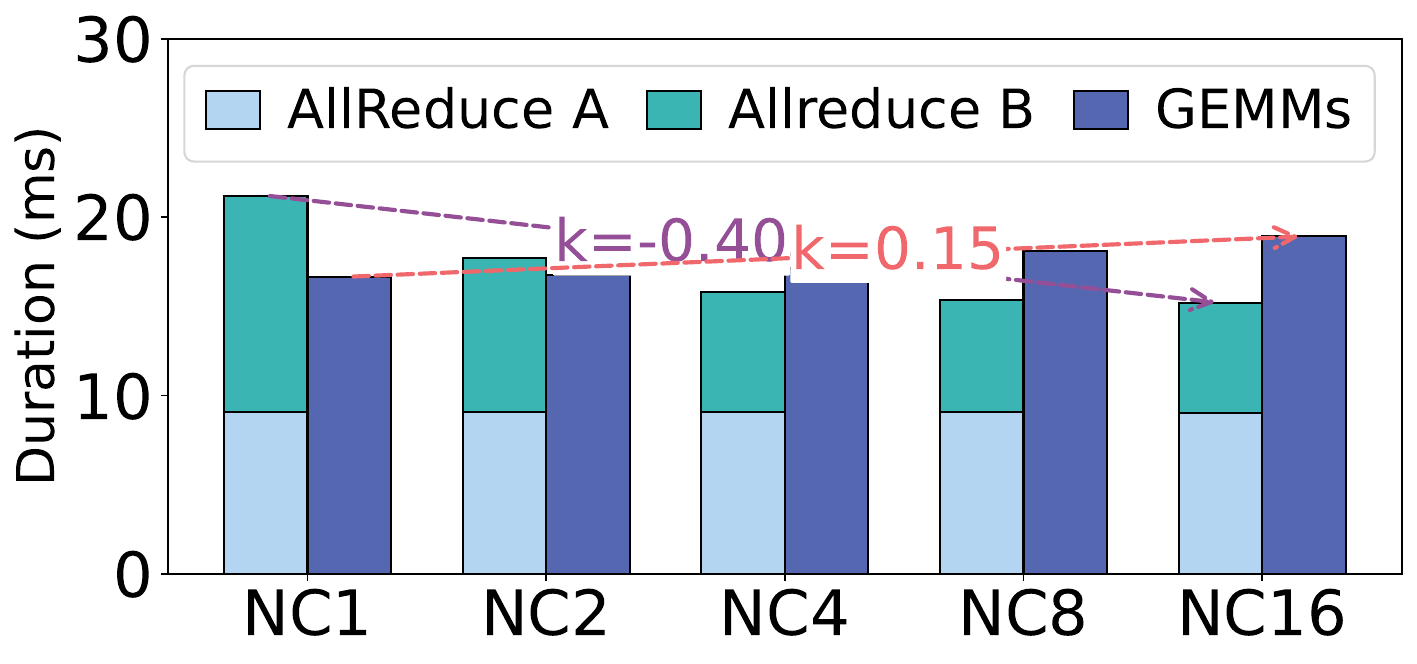}
        \subcaption{tuning AllReduce B only}
        \label{fig:tune.B}
    \end{minipage}
    \caption{Cost differences when tuning different communications in multiple communication scenarios}
    \label{fig:tune_diff_comm}
\end{figure}

As shown in Fig.~\ref{fig:tune_diff_comm}, increasing NC from 1 to 16 affects the two communications differently.
The slope of each arrow indicates the rate of change in computation and communication times.
Optimizing Communication B achieves a larger reduction in communication time with only a slight increase in computation time, resulting in a shorter total execution time.
In contrast, tuning Communication A provides smaller benefits.
These results reveal that optimizing different communications yields distinct trade-offs between communication gains and computational overhead.
The overall performance improvement thus depends on selectively prioritizing the communications that provide higher benefits.
This motivates the introduction of the metric $H$, which quantifies such trade‑offs across concurrent communications.

\noindent\textbf{Metric H.}
We define a priority metric $H_j$ for the j-th communication as Equation \eqref{eq:H}.
$Y$ and $Y'$ denote the computation times before and after tuning, while $x_j^{s_j}$ and $x_j^{s'_j}$ represent the corresponding communication times for the j-th communication.
Where $Y$ and $Y'$ represent the computation time before and after optimization, respectively, and $x_j^{s_j}$ and $x_j^{s'_j}$ represent the communication time corresponding to the two configurations $s_j$ and $s'_j$ for the j-th communication.
\begin{equation}
\label{eq:H}
H_j = \frac{Y'-Y}{x_j^{s_j}-x_j^{s'_j}} 
\end{equation}

Hence, a smaller numerator indicates a lower computational overhead, and a larger denominator implies greater communication improvement.
The metric $H_j$ measures the tuning efficiency, expressed as the communication improvement achieved per unit of added computation cost. A negative denominator implies that the communication is already optimal.

Noted: in multi-computation scenarios, tuning one communication affects multiple overlapped computations, making explicit modeling of each interference infeasible. Since $H_j$ inherently reflects the overall impact of the tuning on all computations, the analysis can be simplified by considering only the aggregate computation time $Y$, significantly reducing complexity.

\subsection{Search Method}
\DontPrintSemicolon
\begin{algorithm}[!htbp]
    \caption{Cost-Effectiveness Algorithm}
    \label{algorithm:cost}
    \SetKwInOut{Input}{Input}
    \SetKwInOut{Output}{Output}
    \Input{Communication config set $S=\{s_1, s_2, ..., s_n\}$}
    \Output{fine-tuned communication config $S'$}
    \BlankLine
    $S' \gets \{\}$\;
    Initialize all $H$ to 0.01\;
    \While{$\exists s \in S$ and $s.done = \text{False} $ } {
        $s_j \gets \arg\min\limits_{s_i \in M, s_i.done = \text{False}} H_i$\;
        $s_j' \gets \text{ResourceEfficientTuning}(s_j)$\;
    \eIf{$s_j'.done$}{
            $S'.\text{append}(s_j)$\;
        }{
            $H_j \gets \frac{Y' - Y}{x_j^{s_j}-x_j^{s'_j}}$\;
        }
    }
    \Return{$S'$}\;
\end{algorithm}

In an overlap, the optimal communication configuration satisfies one of the following conditions: (1) All communications use minimal resources, and their total time cost is less than the computation time. (2) Communications are fully optimized but still take longer than computation, and further increasing communication resources prolongs both communication and computation time. (3) Increasing communication resources reduces communication time but increases computation time until the two are equal—beyond this point, further resource allocation for communication continues to speed up communication while slowing computation even more, degrading overall performance.

\begin{algorithm}[htbp]
    \caption{Resource-Efficient Tuning Algorithm}
    \label{algorithm:resource}
    \SetKwInOut{Input}{Input}
    \SetKwInOut{Output}{Output}
    \Input{Communication config $s_j'$}
    
    \If{$s_j'$ is not initialized}{
        $s_j'.NC,s_j'.NT,s_j'.C\gets NC_{min},NT_{min},C_{min} $ \;
    }
    $x_j^{s_j'}, Y', X' \gets \text{ProfileTime}(s_j')$ \;        
    \eIf{$x_j^{s_j'} - x_j^{s_j} > 0 \text{ or } X' < Y'$}{
        $s_j'.done \gets \text{True}$ \;
    }{
        $lr \gets \frac{x_j^{s_j'} - x_j^{s_j}}{x_j^{s_j'}}$ \;
        $s_j'.NC \gets s_j'.NC + lr$ \;
        $s_j'.NT \gets s_j'.NT + lr$ \;
        $s_j'.C \gets s_j'.C + lr$\;
    }
    \Return $s_j'$
\end{algorithm}

To efficiently identify the optimal configuration that satisfies the boundary conditions, we designed Algorithm~\ref{algorithm:cost} and Algorithm~\ref{algorithm:resource}.

\noindent\textbf{Algorithm~\ref{algorithm:cost}} iteratively optimizes communications within an overlap group. 
In each iteration, it selects the communication with the smallest $H$ from the unfinished set (line 4), then invokes Algorithm~\ref{algorithm:resource} to determine the next configuration for the selected communication (line 5).
If the communication is marked as completed, its configuration is added to the output set (line 7). 
Otherwise, the algorithm updates the corresponding $H$ based on the latest measurements (line 9). 
The process repeats until all communications have been tuned.

\noindent\textbf{Algorithm~\ref{algorithm:resource}} generates the next configuration for a given communication. 
If the communication has not been initialized, NC, NT, and C are set to their minimum values (lines 1-3).
Otherwise, the algorithm checks whether the current configuration satisfies either of the following termination conditions (line 5): (1) the communication itself has reached its optimal performance, or (2) the communication time is less than the computation time. 
If neither condition is met, the algorithm increases the values of NC, NT, and C by a learning rate $lr$ computed from the relative improvement in communication time (lines 8–11).

\begin{figure}[!htbp] 
    \centering
    \includegraphics[width=0.97\linewidth]{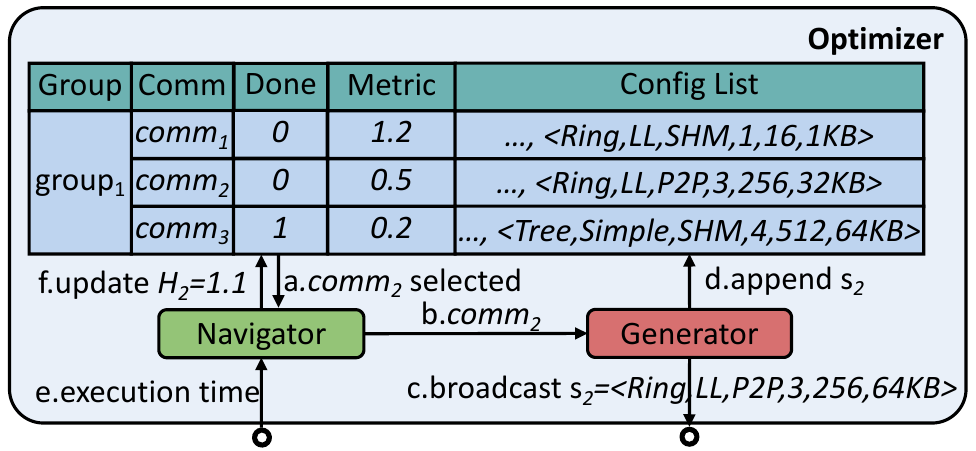} 
    \caption{Workflow of tuning multi-comms in one overlap}
    \label{fig:search.flow} 
\end{figure}

We integrate the proposed algorithms into AutoCCL's divide-and-conquer framework to tune resource-related parameters. 
The overall workflow is illustrated in Fig.~\ref{fig:search.flow}, using an overlap with 3 communication as an example. 
The algorithm begins by selecting the communication with the smallest $H$ among those not yet tuned (\eg, $comm_2$ in step a). 
It then increases the NC, NT, and C for this communication appropriately.
The new configuration $s_2$ is added to $comm_2$'s configuration list (step d) and broadcast to all ranks for execution (step c). 
After measuring the execution time (step e), the $H$ for $comm_2$ is updated (step f). 
This process repeats iteratively until all communications have been tuned.
This approach identifies minimal resource-related parameter that balance communication and computation performance.

\section{Evaluation}
\label{sec:eval}



\begin{figure*}[!h]
    \centering
    \begin{subfigure}[t]{0.92\linewidth}
        \centering
        \includegraphics[width=\linewidth]{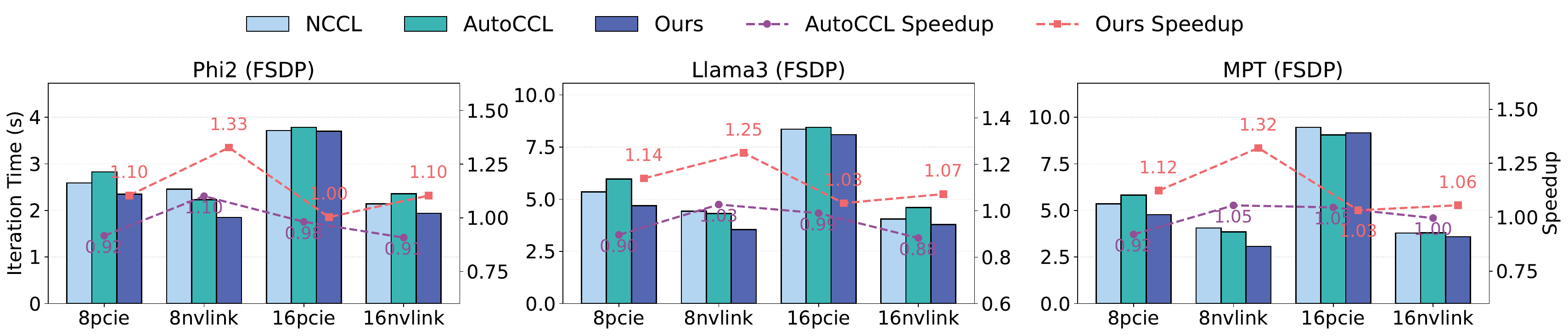}
        \caption{Training iteration time for different models between NCCL, AutoCCL, and Lagom through FSDP}
        \label{fig:eval.end2end.FSDP}
    \end{subfigure}

    \begin{subfigure}[t]{0.95\linewidth}
        \centering
        \includegraphics[width=\linewidth]{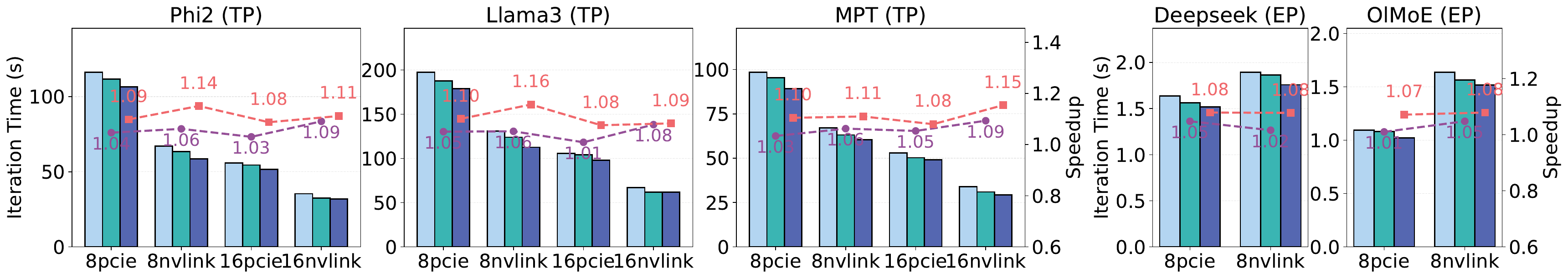}
        \caption{Training iteration time for different models between NCCL, AutoCCL, and Lagom through TP/EP. }
        \label{fig:eval.end2end.TPEP}
    \end{subfigure}
    \caption{
    Overall comparison of iteration time across communication strategies in FSDP (a) and TP/EP (b)
    }
    \label{fig:eval.end2end.overall}
\end{figure*}
\subsection{Experimental Setup}
\label{sec:eval:setup}

\begin{table}[!htbp]
\centering
\caption{DNN model statistics for different parallelism (8-16 A40 GPUs). MBS/GBS denotes Micro/Global Batch Size.}
\label{tab:model.setup}
\resizebox{0.97\columnwidth}{!}{
\begin{tabular}{@{}ccccccc@{}}
\toprule
\textbf{Model} & \textbf{MBS} & \textbf{GBS} & \textbf{TP} & \textbf{DP} & \textbf{EP} & \textbf{FSDP} \\
\midrule
Phi-2-2B & 2 & $2\times[8,16]$ & -- & -- & -- & 8,16\\
Llama-3-8B & 1 & $2\times[8,16]$ & -- & -- & -- & 8,16 \\
MPT-7B & 1 & $2\times[8,16]$ & -- & -- & -- & 8,16 \\
\midrule
Phi-2-2B & 8 & 512 & 8 & 1,2 & 1 & -- \\
Llama-3-8B & 4 & 256 & 8 & 1,2 & 1 & -- \\
MPT-7B & 2 & 256 & 8 & 1,2 & 1 & -- \\
\midrule
DeepSeek-MoE-16B & 2 & 16 & 1 & 1 & 8 & -- \\
OLMoE-1B-7B & 2 & 16 & 1 & 1 & 8 & -- \\
\bottomrule
\end{tabular}
}
\end{table}

\noindent\textbf{Workload and Software}
As shown in Table~\ref{tab:model.setup}, we evaluate our method on a diverse set of large-scale models, including dense models (Phi-2B~\cite{microsoftPhi2Surprising}, Llama-3-8B~\cite{touvron2023llama}, and MPT-7B~\cite{mosaicml2023mpt}) as well as Mixture-of-Experts (MoE) models (DeepSeek-Moe (16B)~\cite{dai2024deepseekmoeultimateexpertspecialization} and OLMoE-1B-7B~\cite{muennighoff2025olmoeopenmixtureofexpertslanguage}). Our implementation is built upon Megatron-LM~\cite{shoeybi2019megatron} and Pytorch 2.3.0, with Data Parallelism (DP), Tensor Parallelism (with DeepSpeed-Domino~\cite{wang2024domino}), Expert Parallelism (with two batch overlapping~\cite{deepep2025}), and FSDP. Across all experiments, the micro-batch size was set to the maximum value that could fit within GPU memory, while the global batch size followed widely-adopted industry standards~\cite{xu2025autoccl}.

\noindent\textbf{Hardware Infrastructure.}
To evaluate the generality of our approach, we use two distinct hardware clusters.
\textbf{Cluster A} consists of 2 nodes, each with 8 NVIDIA A40 GPUs, connected internally via NVLink with full 400 Gbps connectivity and between nodes using 2×400 Gbps InfiniBand.
\textbf{Cluster B} also has 2 nodes with 8 NVIDIA A40 GPUs per node, but uses PCIe 4.0 for intra-node connections and 100 Gbps InfiniBand for inter-node communication.
The software environments are CUDA 12.8 with Driver 570.13. 

\noindent\textbf{Baseline}
We compare our approach against two baselines: NCCL v2.18.3-1 and AutoCCL.
Our implementation is built on NCCL and AutoCCL, and can be integrated non-intrusively into modern DNN training frameworks.

\subsection{End-to-end Performance}
\label{sec:eval:end2end}

Fig.~\ref{fig:eval.end2end.FSDP} presents \pn consistently achieves 1.10-1.33$\times$ performance over NCCL across different clusters and models with FSDP by automatically identifying appropriate configurations that balance resource utilization.
In contrast, AutoCCL's strategy of aggressively tuning communication parameters can lead to worse end-to-end performance than NCCL in computation-bound scenarios since excessive resource allocation to communication exacerbates the computational bottleneck, as predicted by our contention model.
Furthermore, \pn demonstrates greater performance gains on cluster A since NCCL defaults to larger NC values to exploit the available bandwidth when GPUs are connected via NVLink, which significantly increases contention and worsens computational bottlenecks.
Fig.~\ref{fig:eval.end2end.TPEP} shows \pn achieves speedups of 1.08–1.16$\times$ over NCCL in tensor parallelism and 1.07–1.08$\times$ in expert parallelism, demonstrating its general applicability across different parallelism. 
While AutoCCL also achieves notable speedups of 1.03-1.09$\times$, it consistently underperforms \pn since AutoCCL primarily optimizes communication-bounded overlap scenarios, whereas \pn effectively handles diverse bottleneck that occur throughout the training iteration. 
We provide a detailed breakdown of these scenarios in Section~\ref{sec:eval:breakdown}.

\begin{figure}[!htbp]
    \centering
    \begin{minipage}[t]{0.35\linewidth}
        \centering
        \begin{subfigure}[t]{\linewidth}
            \centering
            \includegraphics[width=\linewidth]{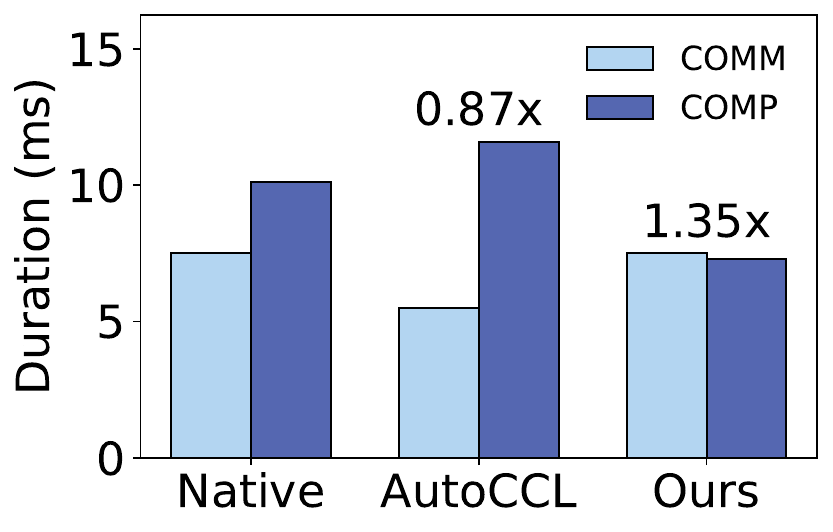}
            \caption{Pattern 1}
            \label{fig:sub_phi1}
        \end{subfigure}
        \\[0.5em]
        \begin{subfigure}[t]{\linewidth}
            \centering
            \includegraphics[width=\linewidth]{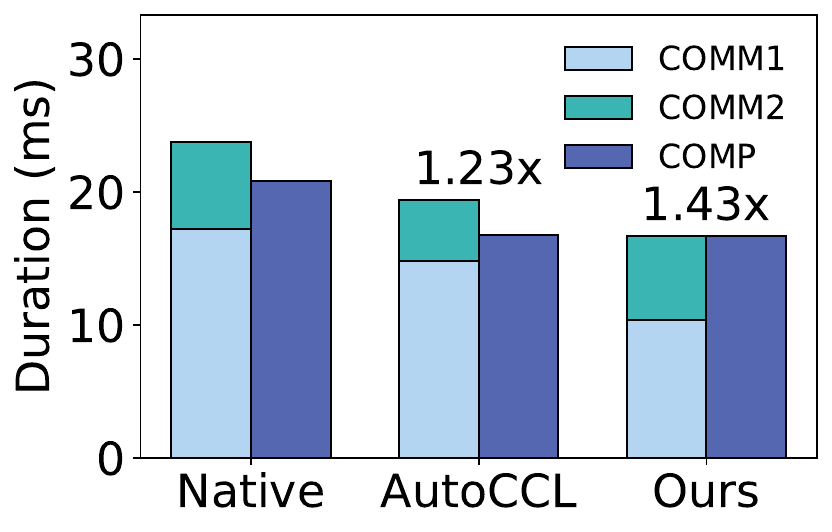}
            \caption{Pattern 2}
            \label{fig:sub_phi2}
        \end{subfigure}
    \end{minipage}
    \hfill
    \begin{minipage}[t]{0.60\linewidth}
        \centering
        \vspace*{-0.05\textheight} 
        \begin{subfigure}[t]{\linewidth}
            \centering
            \includegraphics[width=\linewidth,height=0.62\textheight,keepaspectratio]{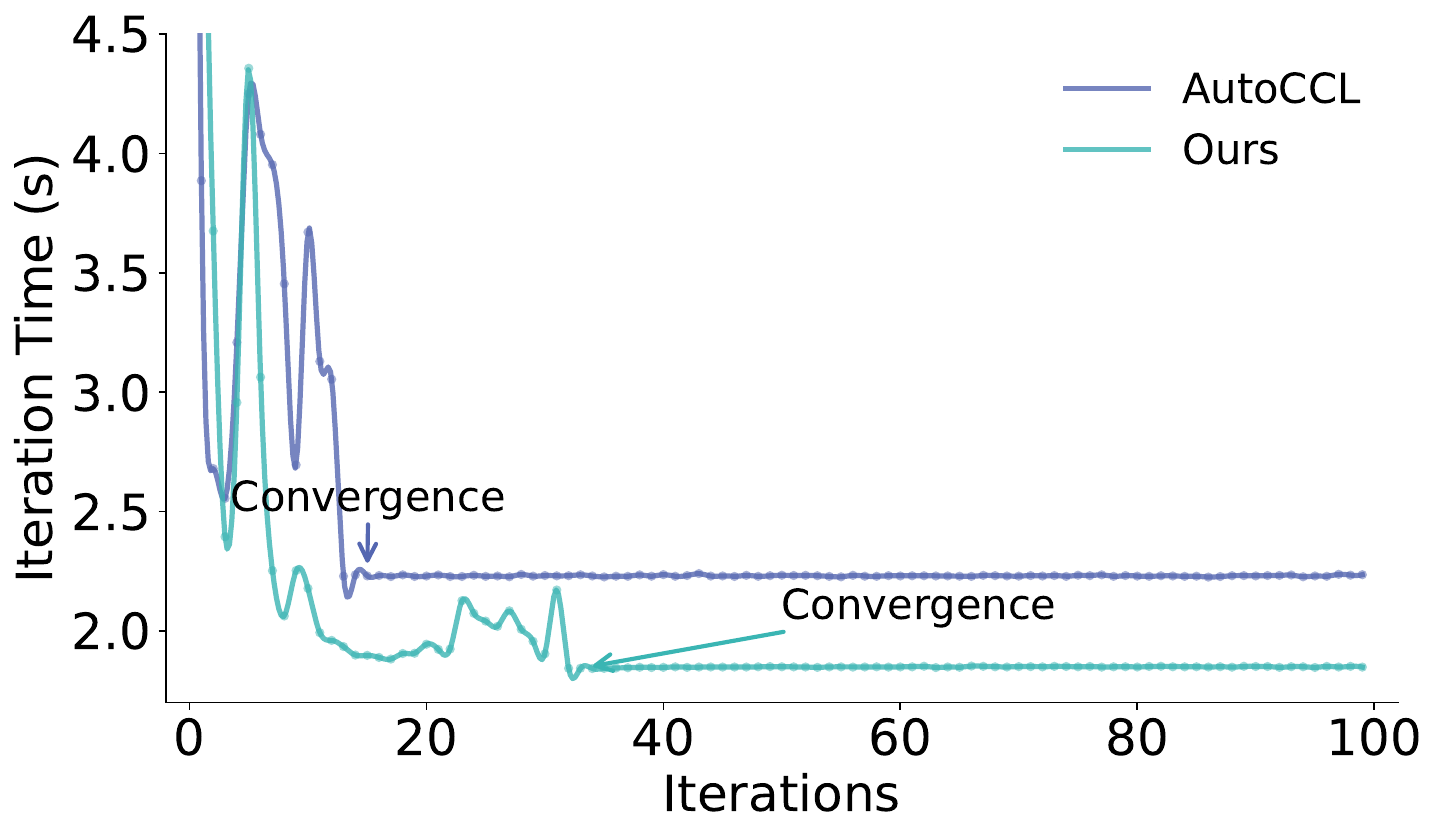}
            \caption{Convergence}
            \label{fig:sub_conv}
        \end{subfigure}
    \end{minipage}
    \caption{Phi2B model breakdown and convergence}
    \label{fig:phi_combined}
\end{figure}

\subsection{Breakdown}
\label{sec:eval:breakdown}
We profile the Phi‑2‑2B model under FSDP on NVLink within a single node and identify two dominant overlap patterns (Fig.~\ref{fig:phi_combined}) covering approximately 90\% of end‑to‑end training time.
As shown in Fig.~\ref{fig:phi_combined}, we analyze the FSDP training of the Phi-2-2B model on a single node in Cluster A. 
In Pattern 1, which is computation-bound, NCCL uses the default configuration of NC=8 and C=2 MB for the \texttt{AllGather}. 
AutoCCL increases the NC to 61 to reduce communication time, but this introduces severe contention and degrades overall throughput to 0.87$\times$. 
In contrast, \pn adopts a configuration of NC=2 and C=684 KB, which slightly reduces communication speed but significantly alleviates contention, yielding an end-to-end speedup of 1.35$\times$.
Pattern 2 involves multiple communications. \pn prioritizes the tuning of the \texttt{ReduceScatter} based on its $H$, adjusting its parameters from (NC=8, C=2 MB) to (NC=4, C=1366 KB), and achieves a 1.43$\times$ speedup.

\subsection{Efficiency of Tuning}
Fig.~\ref{fig:sub_conv} validates that \pn can find the optimal configuration with linear search complexity. 
In an overlap scenario involving two communication tasks, AutoCCL converges within 16 iterations while \pn converges in 33 iterations, showing a ratio of approximately 1:2, which aligns with the predicted linear complexity. 
The resulting convergence overhead is negligible compared to a full training run (over 1 million iterations~\cite{shoeybi2019megatron, jiang2024megascale}), demonstrating the search efficiency and practical utility of \pn.

\section{Conclusion}
\label{sec:conclusion}
In this paper, we present \pn, a novel co-tuning method that automatically and efficiently optimizes communication parameters to speed up the performance of overlap. 
We formalize a cost model for overlap and comprehensively analyze its key challenges under different scenarios. 
To address the exponential search complexity, we introduce a cost-benefit metric and the tuning algorithm that reduces the complexity to linear. 
We implement \pn on top of NCCL and evaluate it on different clusters. 
Results show that \pn achieves 1.07-1.33$\times$ and  1.03-1.27$\times$ performance over both NCCL and AutoCCL across various large-model parallel training.


\bibliography{sections/bibfile}

\end{document}